\documentclass[twocolumn,showpacs,preprintnumbers,amsmath,amssymb]{revtex4}

\usepackage{graphicx}
\usepackage{dcolumn}
\usepackage{bm}

\newcommand{\bq}{\begin{equation}}
\newcommand{\eq}{\end{equation}}
\newcommand{\bqa}{\begin{eqnarray}}
\newcommand{\eqa}{\end{eqnarray}}
\newcommand{\nn}{\nonumber \\}
\newcommand{\ij}{\langle i j \rangle}

\def\be     {\begin{equation}}
\def\ee     {\end{equation}}
\def\bea        {\begin{eqnarray}}
\def\eea        {\end{eqnarray}}
\def\bnn    {\begin{eqnarray*}}
\def\enn    {\end{eqnarray*}}

\begin{document}

\title{SU(2) slave-rotor theory of the attractive Hubbard model}
\author{Ki-Seok Kim}
\affiliation{ School of Physics, Korea Institute for Advanced
Study, Seoul 130-012, Korea }
\date{\today}

\begin{abstract}
Extending the U(1) slave-rotor representation\cite{FG_SRR} of the
repulsive Hubbard model, we propose an SU(2) slave-rotor
decomposition for the attractive Hubbard model, where the SU(2)
slave-rotor variables represent order parameter fluctuations
associated with superconductivity and charge density wave. This
decomposition method allows us to modify the standard Hartree-Fock
mean field theory by incorporating order parameter fluctuations on
an equal footing. Deriving an effective SU(2) slave-rotor action
from the attractive Hubbard model, and analyzing it at the mean
field level, we demonstrate a second order phase transition driven
by softening of the slave-rotor variables.
\end{abstract}

\pacs{71.10.Hf, 74.20.Fg, 71.30.+h, 74.20.-z}

\maketitle

Recently, Florens and Georges proposed a slave-rotor
representation decomposing bare electrons $c_{\sigma}$ into
collective charge excitations $e^{-i\theta}$ and renormalized
electrons $f_{\sigma}$, i.e., $c_{\sigma} =
e^{-i\theta}f_{\sigma}$.\cite{FG_SRR} Applying this representation
to the repulsive Hubbard model, they explained the Mott-Hubbard
transition from a spin liquid Mott insulator to a Fermi liquid
metal in the square lattice. Although the basic scheme of the
rotor representation is quite
appealing,\cite{FG_SRR,Lee_SRR,Kim_SRR} there are several
unsatisfactory points in this approach. First of all, effects of
spin fluctuations are not well described. This is the reason why
the previous slave-rotor theories considered only paramagnetic
phases.\cite{FG_SRR,Lee_SRR,Kim_SRR} Even if only charge
fluctuations are taken into account, the slave-rotor
representation is not complete in the sense that the SU(2)
pseudospin symmetry in the Hubbard model\cite{Pseudospin} is not
reflected in the U(1) slave-rotor representation. Recently, we
extended the U(1) rotor formulation into an SU(2) one for SU(2)
charge fluctuations in the repulsive Hubbard
model.\cite{Kim_metal} In this study we found an anomalous
metallic phase with a pseudogap.

In this paper we apply the SU(2) slave-rotor
representation\cite{Kim_metal} to the attractive Hubbard model.
One main difference from the study of the repulsive Hubbard model
is the presence of nonzero order parameters associated with
superconductivity (SC) and charge density wave (CDW). An important
task is to develop how to incorporate order parameter fluctuations
in the conventional mean field theory such as the BCS scheme. A
standard weak coupling procedure is to integrate out electron
excitations and expand the resulting logarithmic action for order
parameter fluctuations around the mean field ground state, called
the Landau-Ginzburg-Wilson (LGW) expansion. Although this approach
is systematic and firm-based, the procedure to integrate out
gapless electrons near the Fermi surface gives rise to several
uncertainties in the LGW effective theory,\cite{Expansion} thus it
is necessary to treat both electron excitations and order
parameter fluctuations on an equal footing.

In this paper we derive an effective theory from the attractive
Hubbard model, imposing order parameter fluctuations at the mean
field level. Our mean field action consists of two parts. One is a
fermion sector corresponding to a modified Hartree-Fock theory,
and the other a boson part reflecting order parameter
fluctuations, thus allowing us to take into account both electrons
and order parameter excitations on an equal footing. It turns out
that SU(2) rotor variables represent order parameter fluctuations,
and their presence in the effective action admits us to analyze
effects of their fluctuations in the mean field approximation. We
discuss how this mean field scheme modifies the conventional
Hartree-Fock theory.

We consider the attractive Hubbard Hamiltonian \bqa && H = -
t\sum_{\ij\sigma}(c_{i\sigma}^{\dagger}c_{j\sigma} + H.c.) -
\frac{3g}{2}\sum_{i}c_{i\uparrow}^{\dagger}c_{i\uparrow}c_{i\downarrow}^{\dagger}c_{i\downarrow}
, \eqa where $t$ is a hopping integral of electrons, and $g$ a
coupling constant of effective attractions. The interaction term
can be decomposed into pairing and density channels in the
following way \bqa && -
\frac{3g}{2}c_{i\uparrow}^{\dagger}c_{i\uparrow}c_{i\downarrow}^{\dagger}c_{i\downarrow}
= -
\frac{g}{2}c_{i\uparrow}^{\dagger}c_{i\downarrow}^{\dagger}c_{i\downarrow}c_{i\uparrow}
-
gc_{i\uparrow}^{\dagger}c_{i\uparrow}c_{i\downarrow}^{\dagger}c_{i\downarrow}
. \nonumber \eqa Using the identity \bqa
n_{i\uparrow}n_{i\downarrow} = \frac{1}{2} (n_{i\uparrow} +
n_{i\downarrow} - 1)^{2} + \frac{1}{2} (n_{i\uparrow} +
n_{i\downarrow} - 1) \nonumber \eqa with $n_{i\sigma} =
c_{i\sigma}^{\dagger}c_{i\sigma}$, and performing the standard
Hubbard-Stratonovich (HS) transformation for each interaction
channel, we obtain \bqa && Z =
\int{Dc_{i\sigma}D\phi_{i}D\varphi_{i}}e^{-\int_{0}^{\beta}{d\tau}
L} , \nn && L =
\sum_{i\sigma}c_{i\sigma}^{\dagger}(\partial_{\tau} -
\mu)c_{i\sigma} -
t\sum_{\ij\sigma}(c_{i\sigma}^{\dagger}c_{j\sigma} + H.c.) \nn &&
-
\sum_{i}(\phi_{i}c_{i\uparrow}^{\dagger}c_{i\downarrow}^{\dagger}
+ H.c.) + \frac{1}{2g}\sum_{i}|\phi_{i}|^{2} \nn && -
\sum_{i}\varphi_{i}(\sum_{\sigma}c_{i\sigma}^{\dagger}c_{i\sigma}
- 1) + \frac{1}{2g}\sum_{i}\varphi_{i}^{2} . \eqa Here $\phi_{i}$
is an SC order parameter associated with an effective pairing
potential, and $\varphi_{i}$ is a CDW order parameter involved
with an effective density potential. We note that the chemical
potential $\mu = \mu_{b} + g/2$ differs from its bare value
$\mu_{b}$.

One can represent Eq. (2) in terms of a Nambu spinor for
convenience in describing superconductivity.\cite{SC_text} Using
the Nambu spinor $\psi_{i} = \left(
\begin{array}{c} c_{i\uparrow} \\ c_{i\downarrow}^{\dagger}
\end{array} \right)$, we obtain \bqa &&  Z =
\int{D\psi_{i}D\phi_{1i}D\phi_{2i}D\phi_{3i}}e^{-\int_{0}^{\beta}{d\tau}
L} , \nn && L =
\sum_{i}\psi_{i}^{\dagger}(\partial_{\tau}\mathbf{I} -
\mu\tau_{3})\psi_{i} -
t\sum_{\ij}(\psi_{i}^{\dagger}\tau_{3}\psi_{j} + H.c.) \nn && -
\sum_{i}(\phi_{1i}\psi_{i}^{\dagger}\tau_{1}\psi_{i} +
\phi_{2i}\psi_{i}^{\dagger}\tau_{2}\psi_{i} +
\phi_{3i}\psi_{i}^{\dagger}\tau_{3}\psi_{i}) \nn && +
\frac{1}{2g}\sum_{i}(\phi_{1i}^{2} + \phi_{2i}^{2} +
\phi_{3i}^{2}) , \eqa where $\phi_{1i}$ and $\phi_{2i}$ are the
real and imaginary parts of the pairing potential $\phi_{i}$,
i.e., $\phi_{i} = \phi_{1i} - i\phi_{2i}$, and $\varphi_{i}$ is
replaced with $\phi_{3i}$ for a unified notation of SU(2)
symmetry. Introducing a pseudospin variable $\vec{\Omega}_{i} =
(\phi_{1i}, \phi_{2i}, \phi_{3i})$, one can express Eq. (3) in the
following compact form \bqa && Z =
\int{D\psi_{i}D\vec{\Omega}_{i}}e^{-\int_{0}^{\beta}{d\tau} L} ,
\nn && L = \sum_{i}\psi_{i}^{\dagger}(\partial_{\tau}\mathbf{I} -
\mu\tau_{3})\psi_{i} -
t\sum_{\ij}(\psi_{i}^{\dagger}\tau_{3}\psi_{j} + H.c.) \nn && -
\sum_{i}\psi_{i}^{\dagger}(\vec{\Omega}_{i}\cdot\vec{\tau})\psi_{i}
+
\frac{1}{4g}\sum_{i}\mathbf{tr}(\vec{\Omega}_{i}\cdot\vec{\tau})^{2}
. \eqa

Since the above effective Lagrangian is quadratic in electron
excitations, one can formally integrate out the $\psi_{i}$ fields
to obtain an effective Lagrangian of the pseudospin order
parameter by expanding the resulting logarithmic term for the
$\vec{\Omega}_{i}$ fields. It should be noted that the expansion
parameter is $g/D$, where $D$ is an electron bandwidth, thus this
expansion can be justified in $g/D << 1$. Although the weak
coupling condition is satisfied, there are still several
unsatisfactory points in this order parameter action. It is
difficult to justify the LGW expansion in the presence of gapless
electrons because they can cause nonlocal interactions between
order parameters, making it unreliable a conventional treatment in
a local effective action.\cite{Expansion} In this respect we do
not integrate out the fermion excitations in deriving an effective
action. Instead, we treat both electrons and order parameter
excitations on an equal footing, as mentioned before.

For the equal treatment of electrons and order parameters, we
apply a strong coupling approach to this problem, meaning to solve
the interaction term first. Remember that the weak coupling
approach is to solve the kinetic energy term first and treat the
interaction term perturbatively based on the non-interacting
fermion ensemble. Using the identity called the $CP^{1}$
representation\cite{Kim_Kondo} \bqa && \vec{\Omega}_{i}\cdot{\vec
\tau} = m_{i}U_{i}\tau_{3}U^{\dagger}_{i} , ~~~~~~~ U_{i} = \left(
\begin{array}{cc} z_{i\uparrow} & - z_{i\downarrow}^{\dagger} \\
z_{i\downarrow} & z_{i\uparrow}^{\dagger} \end{array} \right) ,
\eqa where $z_{i\sigma}$ in the SU(2) matrix field $U_{i}$ is a
boson field with pseudospin $\sigma$ to satisfy the unimodular
constraint $\sum_{\sigma}|z_{i\sigma}|^{2} = 1$, and performing
the gauge transformation \bqa && \eta_{i\sigma} =
U^{\dagger}_{i\sigma\sigma'}\psi_{i\sigma'} , \eqa one can solve
the coupling term from $-
\sum_{i}m_{i}{\psi}_{i}^{\dagger}(\vec{\Omega}_{i}\cdot\vec{\tau})\psi_{i}$
to $- \sum_{i}m_{i}\eta_{i}^{\dagger}\tau_{3}\eta_{i}$, where
$m_{i}$ is an amplitude of the pseudospin order parameter. We call
$z_{i\sigma}$ and $\eta_{i\sigma}$ a bosonic spinon and a
fermionic chargon, respectively.

In this strong coupling approach we find an interesting physics
that order parameter fluctuations $\vec{\Omega}_{i}$ carrying a
pseudospin quantum number $1$ fractionalize into bosonic spinons
$z_{i\sigma}$ with pseudospin $1/2$ in order to screen out the
pseudospin of an electron due to strong interactions. The
components of the $\eta_{i}$ field are given by $\eta_{i} =
\left(\begin{array}{c} \eta_{i\uparrow}
\\ \eta_{i\downarrow}^{\dagger} \end{array} \right) = \left(\begin{array}{c}
z^{\dagger}_{i\uparrow}c_{i\uparrow}
+  z_{i\downarrow}^{\dagger}c_{i\downarrow}^{\dagger} \\
-z_{i\downarrow}c_{i\uparrow} +
z_{i\uparrow}c_{i\downarrow}^{\dagger}
\end{array} \right)$. Another way to say this
fractionalization is that bare electrons $\psi_{i\sigma}$
fractionalize into bosonic spinons $U_{i\sigma\sigma'}$ and
fermionic chargons $\eta_{i\sigma}$, i.e., $\psi_{i\sigma} =
U_{i\sigma\sigma'}\eta_{i\sigma'}$ owing to strong interactions.

Inserting Eqs.(5) and (6) into Eq. (4), we obtain \bqa &&  Z =
\int{D\eta_{i}DU_{i}}\delta(U_{i}^{\dagger}U_{i} - 1)
e^{-\int_{0}^{\beta}{d\tau} L} , \nn && L =
\sum_{i}\eta_{i}^{\dagger}(\partial_{\tau}\mathbf{I} +
U_{i}^{\dagger}\partial_{\tau}U_{i}-
\mu{U}_{i}^{\dagger}\tau_{3}U_{i})\eta_{i} \nn && -
t\sum_{\ij}(\eta_{i}^{\dagger}U_{i}^{\dagger}\tau_{3}U_{j}\eta_{j}
+ H.c.) \nn && - \sum_{i}m_{i}\eta_{i}^{\dagger}\tau_{3}\eta_{i} +
\frac{1}{4g}\sum_{i}\mathbf{tr}(m_{i}\tau_{3})^{2} . \eqa Note
that the integration measure $\int{D\psi_{i}D\vec{\Omega}_{i}}$ in
Eq. (4) is changed into
$\int{D\eta_{i}DU_{i}}\delta(U_{i}^{\dagger}U_{i} - 1)$ in Eq.
(7). The number of integration variables is $4$ in both cases. One
important point in this expression is that we impute interactions
between electrons and order parameter fluctuations to couplings
between chargons (renormalized electrons) and spinons
(fractionalized pseudospins) in the kinetic energy. A standard way
to treat this nontrivial kinetic energy term is to integrate out
the chargon fields, \bqa && Z =
\int{DU_{i}}\delta(U_{i}^{\dagger}U_{i} - 1)\exp\Bigl[-
\int_{0}^{\beta}{d\tau}\frac{1}{4g}\sum_{i}\mathbf{tr}(m_{i}\tau_{3})^{2}
\nn && + \mathbf{tr}\ln\Bigl(\partial_{\tau}\mathbf{I} -
m_{i}\tau_{3} - \mu{U}_{i}^{\dagger}\tau_{3}U_{i} +
U_{i}^{\dagger}\partial_{\tau}U_{i} -
tU_{i}^{\dagger}\tau_{3}U_{j}\Bigr) \Bigr] , \nn \eqa where
$\mathbf{tr}$ in the logarithm means sum over time, space, spin
and matrix elements. Expanding the logarithmic term for the
bosonic spinons $U_{i\sigma\sigma'}$ ($z_{i\sigma}$), one obtains
an effective action of the spinons. One important difference from
the weak coupling approach is that the expansion parameter is
$D/g$ instead of $g/D$. Unfortunately, this conventional strong
coupling approach has an important defect. Metallic physics
(information of a Fermi surface) of electrons is not introduced in
this expression. Actually, expanding the logarithmic term in the
expansion parameter $D/g$, the resulting effective action is known
to be the O(3) nonlinear $\sigma$ model appropriate to an
insulating magnet, describing competition between SC and
CDW.\cite{Kim_SC_CDW}

Instead of integrating out fermions, we decouple the "interacting"
kinetic energy into the conventional "non-interacting" one via the
HS transformation \bqa && -
t(\eta_{i\alpha}^{\dagger}U_{i\alpha\beta}^{\dagger}\tau_{3\beta\gamma}U_{j\gamma\delta}\eta_{j\delta}
+ H.c.) \nn && \rightarrow
t\Bigl[F_{ij}^{\alpha\delta}E_{ij}^{\dagger\alpha\delta} +
E_{ij}^{\alpha\delta}F_{ij}^{\dagger\alpha\delta} \nn && -
(\eta_{i\alpha}^{\dagger}F_{ij}^{\alpha\delta}\eta_{j\delta} +
E_{ij}^{\dagger\alpha\delta}U_{i\alpha\beta}^{\dagger}\tau_{3\beta\gamma}U_{j\gamma\delta})
- H.c. \Bigr] , \eqa where $E_{ij}$ and $F_{ij}$ are HS matrix
fields associated with hopping parameters of the $\eta_{i}$
fermions and $U_{i}$ bosons, respectively. They are
self-consistently determined from the saddle point equations in
the mean field approximation \bqa && E_{ij}^{\dagger\alpha\delta}
= \langle\eta_{i\alpha}^{\dagger}\eta_{j\delta}\rangle , ~~~~~
F_{ij}^{\alpha\delta} =
\langle{U}_{i\alpha\beta}^{\dagger}\tau_{3\beta\gamma}U_{j\gamma\delta}\rangle
. \eqa We make an ansatz for the hopping matrices as \bqa &&
E_{ij}^{\alpha\delta} = E \tau_{3\alpha\gamma} , ~~~
F_{ij}^{\alpha\delta} = F \tau_{3\alpha\gamma} , \eqa where $E$
and $F$ are amplitudes of the hopping parameters. The reason why
we introduce the $\tau_{3}$ matrix is that the fermion sector
should recover the original electron Lagrangian Eq. (4) as the
slave-rotor representation\cite{Kim_SRR} does.

We also perform a mean field decomposition in the coupling term of
the time part as \bqa && \eta_{i\alpha}^{\dagger}
U_{i\alpha\beta}^{\dagger}\partial_{\tau}U_{i\beta\gamma}
\eta_{i\gamma} \approx \eta_{i\alpha}^{\dagger} \langle
U_{i\alpha\beta}^{\dagger}\partial_{\tau}U_{i\beta\gamma} \rangle
\eta_{i\gamma} \nn && + \langle
\eta_{i\alpha}^{\dagger}\eta_{i\gamma} \rangle
U_{i\alpha\beta}^{\dagger}\partial_{\tau}U_{i\beta\gamma} -
\langle \eta_{i\alpha}^{\dagger}\eta_{i\gamma} \rangle \langle
U_{i\alpha\beta}^{\dagger}\partial_{\tau}U_{i\beta\gamma} \rangle
\nn && \equiv \eta_{i\alpha}^{\dagger} h_{i\alpha\gamma}
\eta_{i\gamma} + l_{i\alpha\gamma}
U_{i\alpha\beta}^{\dagger}\partial_{\tau}U_{i\beta\gamma}  -
h_{i\alpha\gamma} l_{i\alpha\gamma} \eqa with the mean field
ansatz of \bqa && h_{i\alpha\gamma} = \langle
U_{i\alpha\beta}^{\dagger}\partial_{\tau}U_{i\beta\gamma} \rangle
\approx h_{i} \tau_{3\alpha\gamma} , \nn && l_{i\alpha\gamma} =
\langle \eta_{i\alpha}^{\dagger}\eta_{i\gamma} \rangle \approx
l_{i} \tau_{3\alpha\gamma} . \eqa This ansatz is consistent with
Eq. (11). The chemical potential term is decoupled in the mean
field level as \bqa && -
\mu\eta_{i\alpha}^{\dagger}U_{i\alpha\beta}^{\dagger}\tau_{3\beta\gamma}U_{i\gamma\delta}\eta_{i\delta}
\approx - \mu\eta_{i\alpha}^{\dagger} \langle
U_{i\alpha\beta}^{\dagger}\tau_{3\beta\gamma}U_{i\gamma\delta}
\rangle \eta_{i\delta} \nn && - \mu \langle
\eta_{i\alpha}^{\dagger}\eta_{i\delta} \rangle
U_{i\alpha\beta}^{\dagger}\tau_{3\beta\gamma}U_{i\gamma\delta} +
\mu \langle \eta_{i\alpha}^{\dagger}\eta_{i\delta} \rangle \langle
U_{i\alpha\beta}^{\dagger}\tau_{3\beta\gamma}U_{i\gamma\delta}
\rangle \nn && \equiv - \mu\eta_{i\alpha}^{\dagger}
q_{i\alpha\delta} \eta_{i\delta} - \mu l_{i\alpha\delta}
U_{i\alpha\beta}^{\dagger}\tau_{3\beta\gamma}U_{i\gamma\delta} +
\mu l_{i\alpha\delta} q_{i\alpha\delta} \eqa with \bqa &&
q_{i\alpha\delta} = \langle
U_{i\alpha\beta}^{\dagger}\tau_{3\beta\gamma}U_{i\gamma\delta}
\rangle = q_{i}\tau_{3\alpha\delta} . \eqa

Inserting Eqs. (9), (11), (12), (13), (14), and (15) into Eq. (7),
we find an effective Lagrangian for the mean field analysis of the
attractive Hubbard model \bqa && Z =
\int{D\eta_{i}DU_{i}}\delta(U_{i}^{\dagger}U_{i} - 1)
e^{-\int_{0}^{\beta}{d\tau} L} , \nn && L = L_{0} + L_{\eta} +
L_{U} , \nn && L_{0} = 4t\sum_{\ij}EF - 2\sum_{i}h_{i}l_{i} +
2\mu\sum_{i}q_{i}l_{i} + \frac{1}{2g}\sum_{i}m_{i}^{2} , \nn &&
L_{\eta} = \sum_{i}\eta_{i}^{\dagger}(\partial_{\tau}\mathbf{I} -
m_{i}{\tau}_{3} + h_{i}\tau_{3} - \mu{q}_{i}\tau_{3})\eta_{i} \nn
&& - tF\sum_{\ij}(\eta_{i}^{\dagger}\tau_{3}\eta_{j} + H.c.) , \nn
&& L_{U} =
\sum_{i}l_{i}\mathbf{tr}(U_{i}^{\dagger}\partial_{\tau}U_{i}
\tau_{3}) -
tE\sum_{\ij}\mathbf{tr}(U_{i}^{\dagger}\tau_{3}U_{j}\tau_{3} +
H.c.) \nn && -
\mu\sum_{i}l_{i}\mathbf{tr}(U_{i}^{\dagger}\tau_{3}U_{i} \tau_{3})
. \eqa In this effective Lagrangian $l_{i}$, $h_{i}$, $q_{i}$ and
$m_{i}$ are not all independent. One can easily see $2l_{i} =
m_{i}/g$. Introducing $- h_{ri} = h_{i} - \mu q_{i}$ in Eq. (16),
we obtain \bqa && Z =
\int{D\eta_{i}DU_{i}}\delta(U_{i}^{\dagger}U_{i} - 1)
e^{-\int_{0}^{\beta}{d\tau} L} , \nn && L = L_{0} + L_{\eta} +
L_{U} , \nn && L_{0} = 4t\sum_{\ij}EF +
\frac{1}{g}\sum_{i}h_{ri}m_{i} + \frac{1}{2g}\sum_{i}m_{i}^{2} ,
\nn && L_{\eta} =
\sum_{i}\eta_{i}^{\dagger}(\partial_{\tau}\mathbf{I} -
m_{i}{\tau}_{3} - h_{ri}\tau_{3})\eta_{i} -
tF\sum_{\ij}(\eta_{i}^{\dagger}\tau_{3}\eta_{j} + H.c.) , \nn &&
L_{U} =
\frac{1}{2g}\sum_{i}m_{i}\mathbf{tr}(U_{i}^{\dagger}\partial_{\tau}U_{i}
\tau_{3}) -
tE\sum_{\ij}\mathbf{tr}(U_{i}^{\dagger}\tau_{3}U_{j}\tau_{3} +
H.c.) \nn && -
\frac{\mu}{2g}\sum_{i}m_{i}\mathbf{tr}(U_{i}^{\dagger}\tau_{3}U_{i}
\tau_{3}) . \eqa

To explore the consequences of order parameter fluctuations in the
effective rotor action Eq. (17) at the saddle point level, we
express Eq. (17) in terms of the spinons $z_{i\sigma}$ with the
mean field ansatz $m_{i} = (-1)^{i}m$ and $h_{ri} =
(-1)^{i}h_{r}$, \bqa && Z =
\int{D\eta_{i}Dz_{i\sigma}}e^{-\int_{0}^{\beta}{d\tau} L} , \nn &&
L = L_{0} + L_{\eta} + L_{z} , \nn && L_{0} = 2t\sum_{\ij}EF +
\frac{1}{g}\sum_{i}h_{r}m + \frac{1}{2g}\sum_{i}m^{2} , \nn &&
L_{\eta} = \sum_{i}\eta_{i}^{\dagger}(\partial_{\tau}\mathbf{I} -
(-1)^{i}m {\tau}_{3} - (-1)^{i}h_{r} \tau_{3} )\eta_{i} \nn && -
tF\sum_{\ij}(\eta_{i}^{\dagger}\tau_{3}\eta_{j} + H.c.) , \nn &&
L_{z} = \frac{m}{g} \sum_{i\sigma} (-1)^{i}
z_{i\sigma}^{\dagger}\partial_{\tau}z_{i\sigma}  -
tE\sum_{\ij\sigma}( \sigma{z}_{i\sigma}^{\dagger}z_{j\sigma} +
H.c.) \nn && -
\frac{\mu{m}}{g}\sum_{i\sigma}(-1)^{i}\sigma{z}_{i\sigma}^{\dagger}z_{i\sigma}
+ \lambda\sum_{i}(\sum_{\sigma}|z_{i\sigma}|^{2} - 1) , \eqa where
$\lambda$ is a Lagrange multiplier field to impose the unimodular
constraint, replaced with its mean field value. We also replaced
$2E$ with $E$ in the above.

The SU(2) slave-rotor action consists of two parts. One is the
fermion sector $L_{\eta}$, and the other the boson part $L_{z}$.
The fermion action coincides with a conventional mean field
theory, the Hartree-Fock theory except for the renormalization of
the bandwidth $t\rightarrow tF$, if we regard $m + h_r$ as an
effective magnetic field. At zero temperature the fermion sector
thus always remains in the "magnetic" phase with a gap to
quasiparticle excitations due to Fermi-nesting if half filling in
the square lattice is considered. However, it is important to see
that quantum fluctuations of the $z_{i\sigma}$ bosons reduce the
staggered pseudo-magnetization $m$, compared to the Hartree-Fock
magnetization. The boson action, on the other hand, can be
considered as corrections to the fermion mean field theory due to
order parameter fluctuations, not captured in the LGW framework.
It is basically the same as the $CP^{1}$ Lagrangian of the O(3)
nonlinear $\sigma$ model, but modified by the Berry phase term
$\frac{m}{g} \sum_{i\sigma} (-1)^{i}
z_{i\sigma}^{\dagger}\partial_{\tau}z_{i\sigma}$. The presence of
$\sigma = \pm$ in the spinon Lagrangian $L_{z}$, resulting from
the $\tau_{3}$ matrix, also leads to modification of the $CP^{1}$
Lagrangian. If the $\tau_{3}$ matrix is not utilized in Eq. (11),
the hopping term in $L_{z}$ vanishes, thus Eq. (4) cannot be
recovered from Eq. (18) by following its reverse procedure.

In analyzing the effective SU(2) rotor Lagrangian Eq. (18), we
confine our attention to half-filling ($\mu=0$) in the square
lattice for simplicity, and discuss effects of hole doping later.
Let us begin with the fermion sector. The fermion Lagrangian is
well known because its structure is nothing but the BCS theory.
The mean-field conditions for $m$ and $E$ at $T=0$ read \bqa &&
\frac{m}{2g(m+h_{r})} = \sum'_k \frac{1}{E^{\eta}_{k}}, ~~ DE/2 =
\sum'_{k} \frac{F \epsilon_{k}^{\eta{2}}}{E^{\eta}_{k}} . \eqa
Here $\epsilon_{k}^{\eta} = - 2t (\cos{k}_{x} + \cos{k}_{y})$ is
the bare band in the absence of effective exchange splitting
introduced by non-zero $m + h_{r}$, and $E^\eta_k =
\sqrt{(F\epsilon_{k}^{\eta})^2 + (m + h_{r})^2}$ is the chargon
energy with a gap set by $m+h_{r}$. $D = 4t$ is a half of the
bandwidth, and the $k$-sum in both equations is over the reduced
Brillouin zone. One important difference from the conventional
Hartree-Fock scheme is the presence of $h_{r}$, modifying the
first equation.

The spinon Lagrangian is given in the energy-momentum space \bqa
&& L_{z} = \sum'_{\sigma k\nu} z_{\sigma k\nu}^{\dagger} \left(
\lambda + E \sigma\epsilon_k^{z} \right) z_{\sigma k\nu} \nn &&
~~~~ + \sum'_{\sigma k\nu} z_{\sigma k+Q \nu}^{\dagger} \left(
\lambda - E \sigma\epsilon_{k}^{z} \right) z_{\sigma k+Q \nu} \nn
&& ~~~~ + \frac{m}{g} \sum'_{\sigma k\nu} i\nu ( z_{\sigma k+Q
\nu}^{\dagger} z_{\sigma k \nu} + z_{\sigma k \nu}^{\dagger}
z_{\sigma k+Q \nu} ) \nn && ~~~~ - \frac{\mu m}{g}\sum'_{\sigma
k\nu} \sigma ( z_{\sigma k+Q \nu}^{\dagger} z_{\sigma k \nu} +
z_{\sigma k \nu}^{\dagger} z_{\sigma k+Q \nu} ) , \eqa where
$\epsilon_{k}^{z} = -2t(\cos{k}_{x} + \cos{k}_{y})$ is the bare
spinon dispersion with $Q=(\pi,\pi)$. The bosonic $k$-sum is also
over the reduced Brillouin zone. The boson part can be
diagonalized using a pair of operators $(\gamma_{\sigma 1k\nu},
\gamma_{\sigma 2k\nu})$ related to $(z_{\sigma k\nu} , z_{\sigma
k+Q \nu})$ by \bqa z_{\sigma k\nu} &=& \frac{1}{\sqrt{2}} (\cosh
\theta_{\sigma k} - \sinh \theta_{\sigma k}) (\gamma_{\sigma
1k\nu} +\gamma_{\sigma 2k\nu} ) , \nn z_{\sigma k+Q \nu} &=&
\frac{1}{\sqrt{2}} (\cosh \theta_{\sigma k} + \sinh \theta_{\sigma
k} ) (\gamma_{\sigma 1k\nu} -\gamma_{\sigma 2k\nu} ). \eqa After
taking $\cosh 2\theta_{\sigma k} = \lambda/E^z_k$, $\sinh
2\theta_{\sigma k} = E \sigma \epsilon_{k}^{z} /E^z_k$, and $E^z_k
= \sqrt{\lambda^2 - (E \epsilon_{k}^{z})^2}$, one gets \bqa && L_z
= i\frac{m}{g}\sum'_{\sigma k\nu} \nu (\gamma_{\sigma
1k\nu}^{\dagger} \gamma_{1k\nu} - \gamma_{\sigma 2k\nu}^{\dagger}
\gamma_{\sigma 2k\nu} ) \nn && ~~~~ - \frac{\mu m}{g}\sum'_{\sigma
k\nu} \sigma (\gamma_{\sigma 1k\nu}^{\dagger} \gamma_{1k\nu} -
\gamma_{\sigma 2k\nu}^{\dagger} \gamma_{\sigma 2k\nu} ) \nn &&
~~~~ +\sum'_{k\nu} E^z_k (\gamma_{\sigma 1k\nu}^{\dagger}
\gamma_{\sigma 1k\nu} +\gamma_{\sigma 2k\nu}^{\dagger}
\gamma_{\sigma 2k\nu} ) .\eqa The boson spectrum $E^z_k$ is gapped
if $\lambda - E{D} > 0$ while $\lambda - E{D} = 0$ leads to the
condensation of $z_{i\sigma}$.

The mean field equations for $\lambda$, $F$, and $h_{r}$ are
obtained to be \bqa && \frac{m}{g\lambda} = \sum'_k \frac{1}{E^z_k
}, ~~~ DF = \sum'_k \frac{Eg}{ m} \frac{\epsilon_{k}^{z2}}{E^z_k }
\eqa and \bqa && h_{r} =
\sum'_{k}\frac{1}{\beta}\sum_{\nu}\frac{2\nu^2m/g}{E_{k}^{z2} +
\nu^2m^2/g^2} . \eqa An important point in Eq. (24) is that if we
do not introduce an energy cutoff in the frequency integral, the
mean field equation (24) diverges after doing the frequency
integral. However, one need not be surprised at this divergence
because it also happens in the mean field equations for $E$ and
$F$ unless we introduce a momentum cutoff, here the bandwidth $D$.
In this respect it is necessary to introduce an energy cutoff. It
is natural to take the energy cutoff as $m$ because the presence
of $m$ allows this decomposition. When evaluating the frequency
integral, we first divide the integral into two parts, divergent
and divergent-free parts. We calculate the divergent part within
the energy cutoff, but the divergent-free part without the energy
cutoff, i.e., performing the Matsubara summation in the
divergent-free integral. As a result, Eq. (24) reads \bqa && h_{r}
= \frac{2g}{\pi} - \frac{g^2}{m^2}\sum'_{k}E_{k}^{z} . \eqa

To get an idea on the analytical structure of the set of
self-consistent equations obtained above, we first rewrite Eqs.
(19), (23) and (25) as the integration over the energy with a
certain density of states $D(\epsilon)$, and approximate it with a
constant value, $D(\epsilon) = 1/(2D)$, thus $\sum'_{k} =
(1/4D)\int_{-D}^{D}d\epsilon$. The mean-field equations are then
given by \bqa && m + h_{r}
=\frac{FD}{\sinh[\frac{FD}{g}\frac{m}{m+h_{r}}]} , \nn && E =
\frac{\sinh[2FD/g] - 2FD/g}{4\sinh^{2}[FD/g]} , \nn &&
\frac{2mED}{g\lambda} = \sin^{-1}(ED/\lambda) , \nn && F = {4m
ED/g\lambda - \sin [4mED/g\lambda ] \over 8(m/g) \sin^2
[2mED/g\lambda]} , \nn && h_{r} = \frac{2g}{\pi} -
\frac{g^2}{{m}^{2}}\frac{ED\sqrt{\lambda^2-(ED)^{2}} +
\lambda^2\sin^{-1}(ED/\lambda)}{4ED} . \nn \eqa The Bose
condensation occurs at $\lambda_{c} = E_{c}D$, giving rise to
$m_c/g = \pi/4 \approx 0.8$. The critical $(g/D)_{c}$ is
determined from the following equation \bqa &&
\frac{1}{4}\frac{(D/g)_{c}[\sinh(D/g)_{c} -
(D/g)_{c}]}{\sinh^{2}[(1/2)(D/g)_{c}]} \nn && +
\frac{1}{2}\frac{(D/g)_{c}}{\sinh\Bigl[
\frac{(\pi/2)(D/g)_{c}}{\pi/4+2/\pi-\frac{1}{4}\frac{(D/g)_{c}[\sinh(D/g)_{c}
- (D/g)_{c}]}{\sinh^{2}[(1/2)(D/g)_{c}]}}\Bigr]} \nn && =
\frac{\pi}{4} + \frac{2}{\pi} . \eqa Solving this equation
numerically, we find $(g/D)_c \approx 0.3$. This means that the
Bose condensation of spinons appears in $g/D < (g/D)_{c}$ while
they become gapped in $g/D > (g/D)_{c}$.

When the bosonic spinons are condensed, this phase is identified
with an SC state or a CDW phase. We cannot determine which phase
arises because of the SU(2) pseudospin symmetry in the Hubbard
model at half filling. If easy plane anisotropy is introduced, the
resulting state would be an SC phase. When the spinons are gapped,
the resulting phase is an insulating state owing to a gap in the
chargon spectrum at half filling. Remember that chargon
excitations are always gapped at half filling due to
Fermi-nesting. Thus, the SU(2) slave-rotor mean field theory shows
a second order phase transition from an SC state to an insulating
phase at half filling, varying the strength of local attractions.
One may interpret the insulating phase as a preformed-pair state
due to strong phase fluctuations of preformed pairs.

So far, we performed a saddle-point analysis and obtained a
mean-field picture, showing a second order phase transition for
the spinon field $z_{i\sigma}$ associated with order parameter
fluctuations. It is natural to ask the stability of the mean-field
picture against gauge fluctuations $a_{ij}$ that appear in the
phase fluctuations of the hopping parameters, $E_{ij} =
E{e}^{ia_{ij}\tau_{3}}\tau_{3}$ and $F_{ij} =
F{e}^{ia_{ij}\tau_{3}}\tau_{3}$, where $E$ and $F$ are the mean
field values obtained before. It should be noted that the U(1)
pseudospin-gauge field $a_{ij}$ is compact, thus allowing
instanton excitations\cite{Polyakov}. From the seminal work of
Fradkin and Shenker\cite{FS_Instanton} we know that there can be
no phase transition between the Higgs and confinement phases. The
order parameter discriminating the Higgs phase from the
confinement one has not been known yet. In this respect only a
crossover behavior is expected. In the present problem the
phase-coherent state corresponds to the Higgs phase while the
phase-incoherent state coincides with the confinement phase.
Applying Fradkin and Shenker's result to the present problem, we
conclude that the second order phase transition turns into a
crossover between the coherent and incoherent phases. The
pseudospin order parameter, being a gauge-invariant quantity,
remains unaffected by the gauge fluctuation.

One cautious person may ask what the crossover means physically.
Since a systematic method for evaluating the electron spectral
function in the confinement phase is not known, it is difficult to
say any physical statements for the spectral function in fact.
However, recalling the recent study that considerable portions of
coherent quasiparticle spectral weights are transferred to
incoherent backgrounds as the local interactions are
increased,\cite{BD} the crossover in this paper can be understood
by transferring the coherent spectral weights of quasiparticles to
the incoherent backgrounds. Actually, our mean field analysis
coincides with this picture.

Away from half filling, a nonzero chemical potential produces easy
axis anisotropy due to the pseudospin dependence in the chemical
potential term, thus favoring a CDW order. Furthermore, metallic
physics of chargon excitations appears because Fermi-nesting is
destroyed away from half filling. Then, the gapped phase of spinon
excitations may be stable against gauge fluctuations owing to the
presence of gapless fermion excitations.\cite{Hermele_QED3} This
anomalous metallic phase may be related with a pseudogap phase in
the context of high $T_{c}$ superconductivity.

In this paper we developed a mean field theory taking both
electron excitations and order parameter fluctuations on an equal
footing. The effective field theory consists of two parts. One is
a fermion sector corresponding to the conventional Hartree-Fock or
BCS mean field scheme. The other is a boson sector representing
order parameter fluctuations that become fractionalized in the
case of strong interactions, and are expressed as rotor variables.
Softening of rotor fluctuations leads to a second order
transition. Although our mean field analysis was performed in an
insulating phase in the present paper, we believe that the present
slave-rotor formulation would be more useful for studying quantum
phase transitions in itinerant electrons because the presence of
gapless fermion excitations can make the gapped phase of rotor
excitations stable against gauge fluctuations.

Extremely helpful discussions with Prof. J.-H. Han are
appreciated.

\end{document}